\documentclass[aps,prl,twocolumn,showpacs,preprintnumbers,amsmath,amssymb,superscriptaddress]{revtex4-1}

\usepackage{txfonts}
\usepackage{color}
\usepackage{graphicx}
\usepackage{dcolumn}
\usepackage{bm}

\newcommand{\nc}{\newcommand}
\nc{\be}{\begin{eqnarray}}
\nc{\ee}{\end{eqnarray}}
\nc{\bea}{\begin{eqnarray}}
\nc{\eea}{\end{eqnarray}}
\nc{\bean}{\begin{eqnarray*}}
\nc{\eean}{\end{eqnarray*}}
\nc{\mb}{\mbox}
\nc{\rnc}{\renewcommand} 
\nc{\vk}{{\bm k}}
\nc{\vx}{\mb{\bf x}}
\nc{\br}{\mb{\bf r}}
\nc{\bv}{\mb{\bf v}}
\nc{\bp}{\mb{\bf p}}
\nc{\ve}{\mb{\bf e}}
\nc{\vz}{\hat {\mb{\bf z}}}
\nc{\vp}{\mb{\boldmath$p$}}
\nc{\vb}{\mb{\boldmath$b$}}
\nc{\rr}{\mb{\boldmath$r$}}
\nc{\vR}{\mb{\boldmath$R$}}
\nc{\vj}{\mb{\boldmath$j$}}
\nc{\vg}{\mb{\boldmath$g$}}
\nc{\vm}{\mb{\boldmath$m$}}
\nc{\vd}{\mb{\boldmath$d$}}
\nc{\hd}{\mb{\boldmath$\hat{d}$}}
\nc{\vD}{\mb{\boldmath$D$}}
\nc{\vF}{\mb{\boldmath$F$}}
\nc{\vG}{\mb{\boldmath$G$}}
\nc{\vI}{\mb{\boldmath$I$}}
\nc{\vW}{\mb{\boldmath$W$}}
\nc{\x}{\mb{\boldmath$x$}}
\nc{\A}{\mb{\boldmath$A$}}
\nc{\va}{\mb{\boldmath$a$}}
\nc{\vv}{\mb{\boldmath$v$}}
\nc{\vq}{\mb{\boldmath$q$}}
\nc{\vn}{\mb{\boldmath$n$}}
\nc{\vJ}{\mb{\boldmath$J$}}
\nc{\vS}{\mb{\boldmath$S$}}
\nc{\vs}{\mb{\boldmath$\sigma$}}
\nc{\vE}{\mb{\boldmath$E$}}
\nc{\vB}{\mb{\boldmath$B$}}
\nc{\vM}{\mb{\boldmath$M$}}
\nc{\vL}{\mb{\boldmath$L$}}
\nc{\vpsi}{\mb{\boldmath$\psi$}}
\nc{\vphi}{\mb{\boldmath$\varphi$}}
\nc{\Vphi}{\mb{\boldmath$\phi$}}
\nc{\Vomega}{\mb{\boldmath$\Omega$}}
\nc{\ipsi}{\it{\Psi}}
\nc{\vepsilon}{\mb{\boldmath$\epsilon$}}
\nc{\valpha}{\mb{\boldmath$\alpha$}}
\nc{\vgamma}{\mb{\boldmath$\gamma$}}
\nc{\vomega}{\mb{\boldmath$\omega$}}
\nc{\vmu}{\mb{\boldmath$\mu$}}
\nc{\vt}{\mb{\boldmath$\tau$}}
\nc{\vT}{\mb{\boldmath$T$}}
\nc{\vpi}{\mb{\boldmath$\pi$}}
\nc{\nab}{\bm{\nabla}}
\nc{\ov}{\overline}
\nc{\cdott}{\!\cdot\!}
\nc{\cdottt}{\!\!\cdot\!}
\nc{\LL}{\Big{\langle}}
\nc{\RR}{\Big{\rangle}}
\nc{\LR}{\Bigm{|}}
\nc{\vP}{\mb{\boldmath$P$}}
\nc{\nnn}{\nonumber\\}
\rnc{\figurename}{FIG.}

\nc{\psibar}{\overline{\psi}}
\nc{\cbar}{\overline{c}}
\nc{\intx}{\int d^4x}
\nc{\inty}{\int d^4y}
\nc{\intk}{\int \frac{d^4k}{(2\pi)^4}}

\begin{document}

\title{
Voltage-driven magnetization switching and spin pumping in Weyl semimetals
}

\author{Daichi Kurebayashi}
\affiliation{
Institute for Materials Research, Tohoku University, Sendai 980-8577, Japan
}
\author{Kentaro Nomura}
\affiliation{
Institute for Materials Research, Tohoku University, Sendai 980-8577, Japan
}

\date{\today}

\begin{abstract}
We demonstrate electrical magnetization switching and spin pumping in magnetically doped Weyl semimetals.
The Weyl semimetal is a new class of topological semimetals, known to have nontrivial coupling between the charge and the magnetization due to the chiral anomaly.
By solving the Landau-Lifshitz-Gilbert equation for a multilayer structure of a Weyl semimetal, an insulator and a metal whilst taking the charge-magnetization coupling into account, magnetization dynamics is analyzed.
It is shown that the magnetization dynamics can be driven by the electric voltage.
Consequently, switching of the magnetization with a pulsed electric voltage can be achieved, as well as precession motion with an applied oscillating electric voltage.
The effect requires only a short voltage pulse and may therefore be more energetically efficient for us in spintronics devices compared to conventional spin transfer torque switching.

\end{abstract}

%\pacs{73.43.-f, 74.25.fc, 74.90.+n, 74.25.F- }
% PACS, the Physics and Astronomy
                             % Classification Scheme.
%\keywords{Suggested keywords}%Use showkeys class option if keyword
                              %display desired
\maketitle

Controlling magnetization dynamics is one of the challenges for successful applications of spintronic memory, logic, and sensing nanodevices. 
Local magnetic fields and spin-polarized currents have been used to do this \cite{Zutic2004}.
However, there are limitations in applying these schemes: applying local magnetic fields causes difficulties in making scalable systems, whereas employing a (spin polarized) charge current, for spin-transfer torque \cite{Slonczewski1996,Berger1996,Ralph2008,Brataas2012} and spin-orbit torque \cite{Manchon2008a,Miron2011,Liu2012,Mellnik2014}, suffers from Joule heating.
Manipulating the magnetization with gating has also been proposed, however, the high threshold electric voltage is an issue for device applications \cite{Chiba2003,Shiota2012,Wang2011,Kanai2012}.
Scalability, reduced energy dissipation, and a reliable method of controlling magnetization would thus provide meaningful steps for the further development in low energy spintronics devices.

Recently, spintronics phenomena in topological materials have drawn much interest for achieving novel electrical manipulation of  the magnetization. In topological insulator/ferromagnetic insulator heterostructures, magnetization switching \cite{Garate2010,Yokoyama2010,Fan2014}, the control of magnetic textures \cite{Nomura2010}, and the spin-electricity conversion \cite{Shiomi2014} have been examined theoretically and experimentally.

As a new class of topological materials, Weyl semimetals are being researched intensively.
They possess a three-dimensional linear dispersion which is analog of the Weyl fermion in high energy physics \cite{Hosur2013}.
Weyl semimetals can be realized when time-reversal and/or inversion symmetries are broken.
Tantalum arsenide and some other noncentrosymmetric materials are experimentally reported as the inversion symmetry broken Weyl semimetals \cite{Xu2015,Lv,Belopolski2015,Huang2015}.
Although there are many theoretical predictions for the time-reversal symmetry broken Weyl semimetals such as pyrochlore iridates \cite{Wan2011}, the multilayer of the topological and normal insulators \cite{Burkov2011}, and the magnetically doped topological insulators \cite{Liu2013b,Bulmash2014,Zhang2013,Kurebayashi2014}, there have been few materials reported to be time reversal symmetry broken Weyl semimetals \cite{Borisenko2015,Liu2015,Wang2016}.
Since time reversal symmetry broken Weyl semimetals possess both topological and magnetic properties, they might be candidates for new spintronics devices.

The electromagnetic responses in some classes of topologically nontrivial states are described by the Axion term \cite{Qi2011,Zyuzin2012e},
\bea
S_\theta=\int dt d\bm x\left(\frac{e^2}{4\pi^2\hbar c}\right)\theta\bm E\cdot \bm B.
\eea
Particularly in the magnetically doped Weyl semimetals, field theoretical studies addressed that $\theta$ is given by the relation \cite{,Zyuzin2012e},
\bea
\bm \nabla \theta=\frac{x_S S J}{\hbar v_F}\hat{\bm M},
\label{theta}
\eea
where $S$ is spin of the magnetic moments, $x_S$ is a magnetic impurity concentration ratio, $J$ is the exchange coupling constant between a local moment and an itinerant electron, $v_F$ is the Fermi velocity, and $\hat{\bm M}$ is the normalized directional vector of magnetization \cite{Nomura2015}.
Note that Eq.(\ref{theta}) holds even when $\hat{\bm M}$ varies in space.
The charge and current densities are derived by $j^\mu=\delta S_\theta/\delta A_\mu$ as
\bea
\bm j_{\rm AHE}&=&
\sigma_{\rm AHE}\bm \hat{\bm M}\times \bm E
,
\label{Eq-j}
\\
\rho_{\rm AHE}&=&\sigma_{\rm AHE}\bm \hat{\bm M} \cdot \bm B,
\label{Eq-rho}
\eea
where $\sigma_{\rm AHE}$ is the anomalous Hall conductivity defined as $\sigma_{\rm AHE}=\frac{e^2 x_S S J}{2\pi^2 \hbar^2 v_F}$ \cite{Kurebayashi2014}.
The anomalous Hall effect (AHE), Eq. (\ref{Eq-j}), occurs in solids with broken time-reversal symmetry, typically in a ferromagnetic phase, as a consequence of spin-orbit coupling \cite{Nagaosa2010a}.
Equation (\ref{Eq-rho}) states that a charge density is induced by the magnetization in a magnetic field.
We assume in the following that the relation is valid in any magnetic configuration as long as the magnetization varies smoothly with respect to the lattice constants \cite{Nomura2015,Vazifeh2013b}.

In this paper, we study the magnetization dynamics in magnetically doped Weyl semimetals  by solving the Landau-Lifshitz-Gilbert (LLG) equation with effective fields as a consequence of Eq. (\ref{Eq-rho}).
We focus on the case in which the Fermi energy is located at the Weyl points, where the electromagnetic response of the charge density is described by Eq.(\ref{Eq-rho}).
Consequently, we propose a method to switch the magnetization by means of electrical pulses and for generating spin currents by an oscillating electric voltage in the magnetically doped Weyl semimetals.

\begin{figure}[htbp]
\includegraphics[width=1\linewidth]{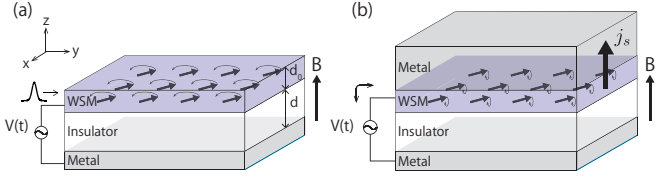}
\caption{(Color online)
Schematic illustration of the multilayer structure of a Weyl semimetal, an insulator, and a metal.
The applied voltage between the Weyl semimetal layer and the bottom metal layer induces magnetization dynamics.
The thickness of the Weyl semimetal and the insulator are denoted as $d_0$ and $d$, respectively.
(a) Voltage pulses switch the magnetization.
(b) Oscillating voltage induces magnetization precession and generates a spin current.
The generated spin current is injected into an adjacent metal layer.
}
\label{setup}
\end{figure}

As a setup, we consider a multilayer device, shown in FIG.\ref{setup}, consisting of a magnetically doped Weyl semimetal, an insulator, and a metal.
An electric voltage is applied between the Weyl semimetal and the metal layers.
A magnetic field is also applied.

We first discuss the energy density functional of the magnetization in the multilayer structure.
In the external magnetic field, the Zeeman contribution to the energy density of local moments is given by
$
E_Z=-g\mu_B \rho_S\hat{\bm M}\cdot\bm B
$
where $\rho_S$ is the density of the magnetic dopants.
Magnetic anisotropy also gives a contribution as
$
E_A=-K\hat M_y^2
$
where $K$ is the anisotropic constant.
Here we consider easy axis anisotropy and take the $y$ axis as the easy axis.

Since the charge couples to the magnetization in the Weyl semimetals, the charge degrees of freedom give two additional contributions to the energy density of the magnetization.
The total number of electrons changes depending on the relative angle between the magnetization and the applied magnetic field.
This induces an additional charging energy density,
\bea
E_C=\frac{\rho_{\rm AHE}^2}{2C}=\frac{1}{2C}\left(\sigma_{\rm AHE}\hat{\bm M}\cdot \bm B\right)^2
\eea
where $C$ is a capacitance per unit volume.
In the presence of an external electric voltage, the electric potential also contributes to the total magnetic energy as
\bea
E_V=V \rho_{\rm AHE}=V\left(\sigma_{\rm AHE}\hat{\bm M}\cdot \bm B\right)
\eea
where $V$ is the applied voltage.
Therefore the total magnetic energy density is given by
\bea
\nonumber E_{\rm total}&=&-g\mu_B \rho_S\hat{\bm M}\cdot\bm B-K\hat M_y^2\\
&& +\frac{1}{2C}\left(\sigma_{\rm AHE}\hat{\bm M}\cdot \bm B\right)^2+V \left(\sigma_{\rm AHE}\hat{\bm M}\cdot \bm B\right).
\label{E_mag}
\eea

To minimize the charging energy, $E_C$, the magnetization lies in the plane perpendicular to the magnetic field, decreasing the induced charge, Eq. (\ref{Eq-rho}).
The Zeeman energy density of the local moments, $E_Z$, favors the magnetization parallel to the magnetic field.
The electic voltage contribution, $E_V$, favors the magnetization direction parallel or antiparallel to the magnetic field depending on the sign of $V$.

\begin{figure}[htbp]
\includegraphics[width=1\linewidth]{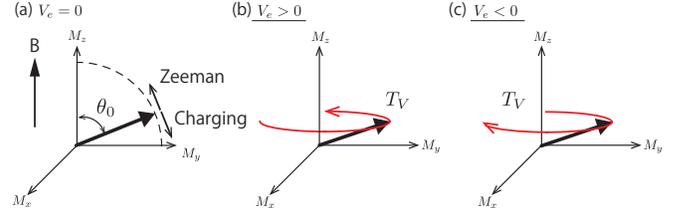}
\caption{(Color online)
Schematic illustration of the spin torque induced by the electric voltage.
(a) The equilibrium magnetization direction determined by the balance of the charging energy and the Zeeman energy.
(b),(c) Magnetization precession driven by the applied voltage.
The direction of the spin torque changes depending on the sign of the voltage.}
\label{torque}
\end{figure}

Let us consider the equilibrium state of the magnetization without the electric voltage.
We take the external magnetic field as $\bm B=B\hat{\bm z}$.
As shown in FIG. \ref{torque} (a) the equilibrium magnetization direction is obtained by minimizing $E_C+E_Z+E_A$ as
\bea
\theta_0 = \cos^{-1} \left(-\frac{Bg\mu_B\rho_S}{\frac{1}{C}\sigma_{\rm AHE}^2B^2+K}\right)
\label{equib_ang}
\eea
when $\frac{1}{C}\sigma_{\rm AHE}^2B^2+K>Bg\mu_B\rho_S$, and $\theta_0=0$ when $\frac{1}{C}\sigma_{\rm AHE}^2B^2+K<Bg\mu_B\rho_S$, where $\theta_0$ is the stabilized zenith angle of the magnetization.
The in-plane component is always $+\hat{\bm y}$ or $-\hat{\bm y}$ direction because of magnetic anisotropy.
Therefore, the magnetization direction can be tilted away from the magnetic field by the charging energy as a consequence of coupling between the magnetization and the charge.
The charging energy becomes dominant in the strong magnetic field regime.
The dominance of energies is also controlled by changing the capacitance.
In the multilayer structure shown in FIG. \ref{setup}, the capacitance is estimated as
$
C=\frac{\varepsilon}{d\ d_0}
$
where $\varepsilon$ is a dielectric constant of the insulator, $d$ and $d_0$ are the thickness of the insulator and the Weyl semimetal.

Now we discuss dynamics of the magnetization in the Weyl semimetal.
The magnetization dynamics is described by the LLG equation,
\bea
\frac{d\hat{\bm M}}{dt}=-\gamma\hat{\bm M}\times \bm B_{\rm eff}+\alpha\hat{\bm M}\times\frac{d\hat{\bm M}}{dt},
\label{LLG}
\eea
where $\bm B_{\rm eff}$ is an effective magnetic field obtained by taking variational of the total energy density by magnetization, 
\bea
\bm B_{\rm eff}=\frac{1}{\hbar\gamma}\frac{\delta E_{\rm total}}{\rho_S\delta \hat{\bm M}},
\eea
$\gamma$ is the gyromagnetic ratio, and $\alpha$ is the Gilbert damping constant.
$E_{\rm total}$ is the total energy density of magnetization introduced in Eq. (\ref{E_mag}).
The charging energy $E_C$ and the potential term $E_V$ give additional contributions to the LLG equation because the charge degrees of freedom couple to the magnetic degrees of freedom \cite{Nomura2015}.
The contributions are described in terms of the effective field, $\bm B_{\rm eff}=\bm B+\bm B_{A}+\bm B_{C}+\bm B_{V}$, as
\bea
\bm B_{C}&=&\frac{\partial E_C}{\hbar\gamma\rho_S \partial \hat{\bm M}}=\frac{\sigma_{\rm AHE}}{\hbar\gamma\rho_S C}\left(\sigma_{\rm AHE}\hat{\bm M}\cdot\bm B\right)\bm B \label{B_c}\\
\bm B_{V}&=&\frac{\partial E_V}{\hbar\gamma\rho_S \partial \hat{\bm M}}=\frac{V\sigma_{\rm AHE}}{\hbar\gamma\rho_S}\bm B.\label{B_v}
\eea
Then spin torque associated with the charging and potential contributions are obtained as $\bm T_C=\hbar\gamma\bm B_C\times\hat{\bm M}$ and $\bm T_V=\hbar\gamma\bm B_V\times\hat{\bm M}$.
The anisotropy energy gives a contribution which is independent of the magnetic field as
$
\bm B_{A}=\frac{\partial E_A}{\hbar\gamma\rho_S \partial \hat{\bm M}}=2\frac{K}{\hbar\gamma\rho_S}M_y\hat{\bm y}.
$
By solving the LLG equation with these additional contributions, the time evolution of magnetization can be computed.

\begin{figure}[tbp]
\includegraphics[width=1\linewidth]{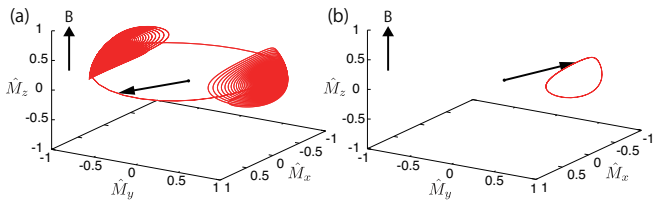}
\caption{(Color online)
(a) Magnetization trajectory with the pulsed voltage.
It shows that the magnetization changes the sign from $+y$ to $-y$.
(b) Magnetization trajectory driven by the oscillating voltage.
It shows that the magnetization precesses around the equilibrium axis.
}
\label{trajectory}
\end{figure}

In our calculation, we choose the parameters as $\sigma_{\rm AHE}=336.3\Omega^{-1}\rm cm^{-1}$, $\rho_S=1.3\times 10^{20}\rm {cm}^{-3}$, $\alpha=0.01$, $\epsilon/\epsilon_0=9.7$ corresponding to silicon carbide where $\epsilon_0$ is the electric constant, $2K/{\hbar \gamma \rho_S}=0.1 \rm T$, $B=0.1 \rm T$, and $d\cdot d_0 = 5.0\times 10^{-15} \rm m^2$.
With these parameters, the charging energy is larger than the Zeeman energy.
Thus the equilibrium magnetization angle is not parallel to the external magnetic field as discussed in Eq.(\ref{equib_ang}).
We examine magnetization dynamics for pulsed and oscillating electric potentials.
Typical numerical results of the magnetization trajectory for each input voltages are shown in FIG. \ref{trajectory}.
It indicates that the magnetization changes its direction between $+\hat{\bm y}$ and $-\hat{\bm y}$ for the pulsive voltage, FIG. \ref{trajectory} (a), and precesses for the oscillating voltage, FIG. \ref{trajectory} (b).
We will explain more detail in the following.

\begin{figure*}[htbp]
\includegraphics[width=1\linewidth]{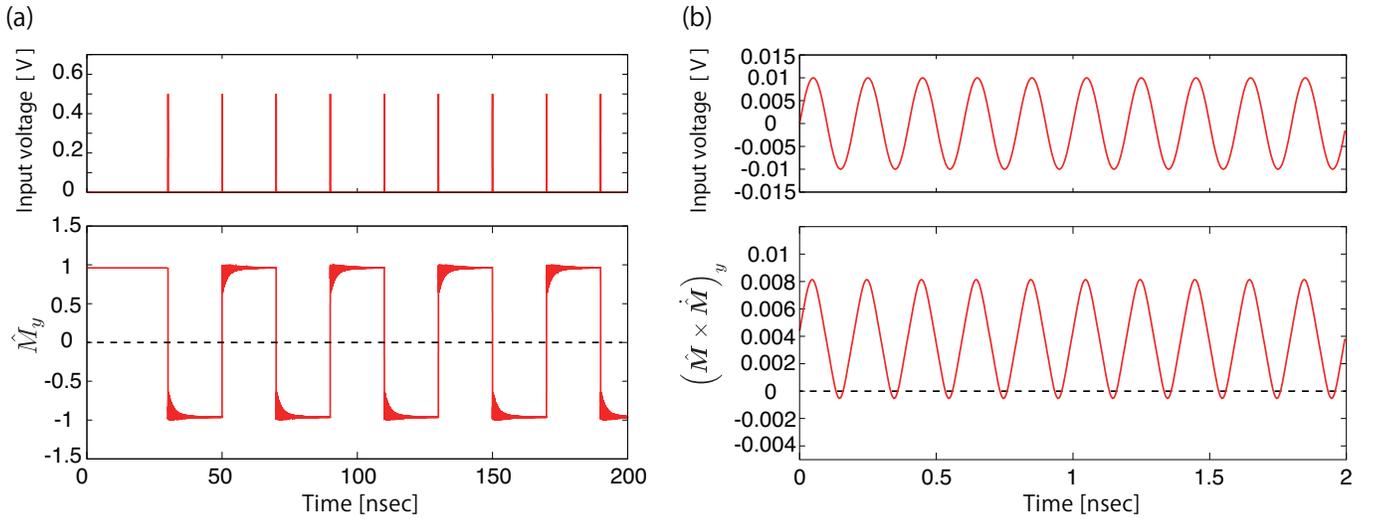}
\caption{(Color online)
(a) The time evolution of the $y$ component of the magnetization with the pulsed voltage.
The numerical calculation is done with the parameters, $V_0=0.1 \rm V$ and $\delta t = 2 \rm nsec$.
Lower figure shows spin switching which $M_y$ changes its sign following the pulsed voltage inputs.
(b) The time evolution of the $y$ component of $\hat{\bm M}\times \dot{\hat{\bm M}}$ with the oscillating voltage.
The parameters are taken as $V_0=0.1 \rm V$ and $\omega = 5 \rm GHz$.
}
\label{results}
\end{figure*}

Let us start with the pulsed electric voltage.
When the electric voltage is absent, $V=0$, the magnetization relaxes to the direction determined by Eq.(\ref{equib_ang}).
In this case, the Zeeman energy and the charging energy counterbalance each other as shown in FIG.\ref{torque} (a).
When the electric voltage $V$ is turned on, $B_V$ is generated and then the spin torque $T_V$ is induced.
The direction of the torque depends on the sign of the applied voltage, $V$, as shown in FIG.\ref{torque} (b) and (c).
When the voltage $V$ is large enough to overcome the anisotropy, the magnetization changes its direction from $+\hat{\bm y}$ to $-\hat{\bm y}$, or vice versa.
We numerically examine the LLG equation for the case of pulsed voltage input.
For each pulse, we use the Gaussian function, $V=V_0 \exp\left[-t^2/2\delta t^2\right]$ where $V_0$ is the amplitude and $\delta t$ is the width of the pulse.
As shown in FIG.\ref{results} (a), the $y$ component of magnetization can be repeatedly reversed by electric pulses.

To compare with a current-induced magnetization dynamics such as the spin-transfer torque \cite{Slonczewski1996,Berger1996,Ralph2008,Brataas2012} and the spin-orbit torque \cite{Manchon2008a,Miron2011,Liu2012,Mellnik2014}, there is no threshold current for the magnetization switching with this mechanism, thus it is expected to be energetically more efficient.
The magnetization control using an electric field has been also proposed and performed experimentally in a magnetic tunnel junction structure comprising a ferromagnetic metal, FeCo, and a MgO barrier \cite{Shiota2012}.
In this structure, the spin torque is generated as a consequence of a change in the perpendicular magnetic anisotropy by the electric voltage.
However a strong electric field is required to produce a large enough torque for magnetization switching with this mechanism, limiting the practical application for nonoscale devices, because the torque is a secondary effect from a changing in the Fermi surface anisotropy.
By contrast, there is the direct coupling between the magnetization and the charge in the Weyl semimetal, Eq. (\ref{Eq-rho}), our method only requires the electric voltage larger than the easy axis anisotropy which can be tuned experimentally.
Therefore the structure with the Weyl semimetal in FIG. \ref{setup} (a) might be suitable for a practical device application.

Next we consider the magnetization dynamics under the oscillating electric voltage.
In addition to controlling the direction of the magnetization, generation of spin currents is one of central issues in the field of spintronics.
Spin pumping is a well established method of generating spin currents, which allows the transfer of the spin angular momentum from magnetization precession motion in a ferromagnet to the conduction electron spin \cite{Tserkovnyak2002,Tserkovnyak2005}.
To induce precession of the magnetization a microwave is irradiated in addition to the static magnetic field tuned at the ferromagnetic resonant condition.
In the following we propose an alternative method to induce precession motion of the localized magnetization in the Weyl semimetal.
In stead of microwave irradiation, we introduce an oscillating voltage under the condition which the applied voltage is smaller than the anisotropic energy.
Since the voltage-induced torque does not overcome the anisotropy torque, it does not lead to reversal but precession of the magnetization about the equilibrium axis given by Eq. (8). 
Figure 3(b) shows a typical trajectory of magnetization precession under an oscillating electric voltage $V = V_0 \sin(\omega t)$ where $\omega$ is the frequency. 

Here we consider a metal attached to the Weyl semimetal as depicted in Fig. 1(b).
In a model of spin pumping the DC component of the spin current density in the adjacent metal layer by the precessing magnetization is expressed as
\bea
\bm j_S = 
\frac{\omega}{2\pi}\int_0^{2\pi/\omega}dt\
g_{\downarrow\uparrow} \hat{\bm M}\times\dot{\hat{\bm M}}
\eea
where $ g_{\downarrow\uparrow}$ is the real part of the spin mixing conductance \cite{Tserkovnyak2002,Tserkovnyak2005} at the interface between the adjacent metal and the Weyl semimetal.
Experimentally the spin current density ${\bm j}_s$ can be detected as a voltage signal via the inverse spin Hall effect in the metal layer \cite{Saitoh2006}.
We compute $\hat{\bm M}\times\dot{\hat{\bm M}}$, and its $y$ component is shown in FIG.\ref{results} (b).
The result shows an oscillating behavior in $(\hat{\bm M}\times\dot{\hat{\bm M}})_y$ with a same frequency as the input electric voltage.
This suggests that an AC spin current is induced by the electric voltage.
There is also DC bias in $(\hat{\bm M}\times\dot{\hat{\bm M}})_y$.
In the multilayer structure, FIG. \ref{setup} (b), the generated spin current is injected into the top metal layer.
An important point here is that, in the Weyl semimetals, magnetization precession motion is induced by the oscillating voltage and the spin pumping is expected.

In conclusion, we have analyzed magnetization dynamics in a magnetically doped Weyl semimetal.
By solving the LLG equation, we found that magnetization dynamics is drastically modified without electric currents due to coupling between the magnetization and the charge density. 
As a result, switching motion of the magnetization is induced by a pulsed electric voltage.
In addition, magnetization precession is induced by an oscillating electric voltage, generating the spin currents.
These electrical manipulations of the magnetization without currents are indispensable for low energy consumption devices, so that the Weyl semimetal might be the candidate of the next generation spintronics material.

D. K. is supported by a JPSJ Research Fellowship for Young Scientists.
This work was supported by Grants-in-Aid for Scientific Research (Nos. 15H05854 and 26400308) from the Ministry of Education, Culture, Sports, Science and Technology, Japan (MEXT).

%--------------------------------------------

\end{document}